\documentclass[aps,prl,twocolumn,floatfix,altaffilletter,superscriptaddress,preprintnumbers, tightenlines,showpacs,showkeys,nofootinbib,notoccite]{revtex4-1}
\usepackage[utf8]{inputenc}
\usepackage[colorlinks=true,citecolor=blue,linkcolor=blue]{hyperref}
\usepackage[normalem]{ulem}
\usepackage{amsmath,amssymb, mathrsfs,esvect}
\usepackage{epsfig}
\usepackage{graphicx}               % Standard graphics package
\usepackage{url}
\usepackage{color}
\usepackage{slashed}
\usepackage{multirow}
\usepackage{placeins}
\usepackage[dvipsnames]{xcolor}
\usepackage{epstopdf}
\usepackage{soul}
\usepackage{tikz}
\usepackage[capitalise, english]{cleveref}
\usepackage{siunitx}
\usepackage{xspace}
\usepackage{booktabs}
\usetikzlibrary{trees}
\usetikzlibrary{decorations.pathmorphing}
\usetikzlibrary{decorations.markings}
\newcommand\myshade{80}
\colorlet{mylinkcolor}{Blue}
\colorlet{mycitecolor}{Red}
\colorlet{myurlcolor}{violet}

\newcommand{\beq}{\begin{eqnarray}}
\newcommand{\eeq}{\end{eqnarray}}

\hypersetup{
	linkcolor  = mylinkcolor!\myshade!black,
	citecolor  = mycitecolor!\myshade!black,
	urlcolor   = myurlcolor!\myshade!black,
	colorlinks = true
}
\allowdisplaybreaks
\begin{document}

%=============================================================================
\title{Is the Turner Window Open?  Seeking Closure \\with Resonant Absorption of Galactic Axions in NaI Dark Matter Detectors}
%\title{Did We Leave the Turner Window Open? \\ \textcolor{red}{Closing} with Resonant Capture of Galactic Axions in NaI Dark Matter Detectors}
\author{W. C. Haxton}
\email{haxton@berkeley.edu}
\affiliation{Department of Physics, University of California, Berkeley, CA 94720, USA}
\affiliation{Institute for Nuclear Theory, University of Washington, Seattle, WA 98195, USA}
\affiliation{Lawrence Berkeley National Laboratory, Berkeley, CA 94720, USA}
\author{Xing Liu} 
\email{xingyzt@berkeley.edu}
\affiliation{Department of Physics, University of California, Berkeley, CA 94720, USA}
\author{Anupam Ray}
\email{anupam.ray@queensu.ca}
\affiliation{The Arthur B. McDonald Canadian Astroparticle Physics Research Institute, Department of Physics, Engineering Physics, and Astronomy, Queen’s University, Kingston, ON, K7L3N6, Canada}
\affiliation{Perimeter Institute for Theoretical Physics, Waterloo, ON, N2L2Y5, Canada}
\author{Evan Rule} 
\email{erule@lanl.gov}
\affiliation{Theoretical Division, Los Alamos National Laboratory, Los Alamos, NM 87545, USA}

\date{\today}
%%%%%%%%%%%%%%%%%%%%%%%%%%%%%%%%%%%%%%%%%%%%%%%%%%%%%%%%%%%%%%%%%%%
\begin{abstract}
Motivated by the DAMA/LIBRA annual modulation signal, the dark matter community has invested heavily in ultra-clean underground NaI detectors to search for light WIMPs.  We point out a new target of opportunity for these detectors --- axions produced by the carbon-burning stars within our galaxy.  These stars synthesize large quantities of $^{23}$Na, keeping it at temperatures $\sim 10^9$K for periods up to tens of thousands of years.  Under these conditions, $^{23}$Na radiates 440~keV axions through repeated photo-excitation and axio-deexcitation of its first excited state. Upon reaching a NaI detector, the process is reversed: the axion is resonantly absorbed, producing a 440 keV deexcitation photon. NaI thus serves as both $\gamma$ source and $\gamma$ detector. We find that existing NaI detectors can probe axion-nucleon couplings $|g_{aNN}^\mathrm{eff~^{23}Na}| \approx g_{app} \sim 10^{-6}$--$10^{-2}$, including QCD axions with $m_a \gtrsim 10$ eV.  While there are several astrophysical constraints on axions with these couplings, our re-examination of these bounds shows that substantial gaps remain, providing strong motivation for the proposed searches.
\end{abstract}

\maketitle
\preprint{N3AS-26-005, INT-26-008, LA-UR-26-22066}
%%%%%%%%%%%%%%%%%%%%%%%%%%%%%%%%%%%%%%%%%%%%%%%%%%%%%%%

The DAMA/LIBRA collaboration \cite{Bernabei:2003za,bernabei2021darkmatterdamalibraperspectives} 
has identified an annual modulation in its data, attributing it to the effects
of the Earth's orbital motion on the WIMP velocity distribution and interaction rate.
The signal is consistent with a light WIMP depositing energies of 2--6 keV through
elastic scattering.  The field has responded by investing heavily in other NaI(Tl) detectors to check this claim, including ANAIS \cite{ntnl-zrn9}, COSINE \cite{yu2025limitswimpdarkmatter}, KIMS \cite{kim2015statuskimsnaiexperiment}, NAIAD \cite{Alner_2005}, NEON \cite{PhysRevLett.134.021802} and SABRE \cite{Milligan_2025}.
Here we point out that these arrays might have another important application,
the detection of galactic axions through resonant absorption, producing a signal
at 440 keV.
\vspace*{.2cm}\\
\noindent
\textbf{{The axion source:}} Some time ago, it was pointed out \cite{PhysRevLett.66.2557} that stars can produce axions \cite{PhysRevLett.38.1440,
PhysRevLett.40.223,PhysRevLett.40.279,Caputo:2024oqc} as
thermally broadened lines through a mechanism in which certain nuclei act
as ``pumps". The mechanism operates through the repeated photo-excitation and axio-deexcitation of a nuclear excited state.  For the mechanism to be effective, the abundance of the relevant isotope must be significant, the gap to the excited state cannot be too much greater than the star's core temperature $T_c$, and the transition probability for axion emission must
be favorable.  If  these conditions are met, significant axion emission will continue throughout the period in which thermal conditions produce a
substantial Boltzmann occupation of the excited state.

These requirements are restrictive, as the odd-$A$ isotope acting as the pump must have a significant abundance, a modest gap, and a strong magnetic transition probability. Stellar compositions, however, are dominated by H and by $\alpha$-stable fusion products such as $^4$He, $^{12}$C, and $^{16}$O with large gaps and E2 transition probabilities, making these isotopes ineffective axion emitters.  
In contrast, odd-$A$ abundances are typically very low, as many such
isotopes burn at high temperatures.
Taking into account the gap and transition strength, 
$^{57}$Fe and $^{23}$Na were identified in \cite{PhysRevLett.66.2557} as the most effective axion emitters
at low and high stellar core temperatures, respectively.   

Over the years, significant attention has been given to axion production by $^{57}$Fe. The 14.4 keV M1 transition, 
familiar from recoilless M{\"o}ssbauer spectroscopy, produces axions most effectively
at $T \sim 10^8$K, a temperature
typical of the cores of red giants.  The Boltzmann occupation of the 14.4 keV $^{57}$Fe excited state is then $\approx 0.27$. In \cite {PhysRevLett.66.2557}, red giant cooling was used to probe the axion-nucleon effective coupling $g^\mathrm{eff~^{57}Fe}_{aNN} \equiv g_{ann}+0.088 g_{app} \approx g_{ann}$ at sensitivities $\gtrsim 10^{-7}$.

Direct searches for these axions have also been performed, using
the Sun as a source: its proximity compensates for its cool
average core temperature, $T_c \approx 1.3 \times 10^7$K. Searches for solar $^{57}$Fe axions have been carried out by
CAST (CERN Axion Solar Telescope), using $a \rightarrow \gamma$ conversion in the 9T CAST magnet~\cite{CASTcollaboration_2009}, by CUORE \cite{CUORE:2012ymr} and Xenon1T \cite{XENON:2020rca}, using the axio-electric effect, and by the authors of \cite{PhysRevLett.75.3222,NAMBA2007398,Derbin,KRCMAR199838},
using the resonant absorption process of present interest.

No searches have been done for $^{23}$Na axions because no strong source of these axions was known.  However, it was recently shown that the
the galaxy's massive stars are such a source~\cite{HLMR}. Unlike $^{57}$Fe, where the isotope
is primordial, incorporated into the star at the time of formation, the $^{23}$Na
is produced {\it in situ} during carbon-burning by the reaction
\begin{equation}
{^{12}\mathrm{C}}+{^{12}\mathrm{C}} \rightarrow {^{24}\mathrm{Mg}}^* \rightarrow  {^{23}\mathrm{Na}}+p  + 2.24\,\,\textrm{MeV}~.
\end{equation}

The contributing stars, with masses $\gtrsim 7.5M_\odot$, are the progenitors of ONeMg white dwarfs (WDs) and electron-capture and core-collapse supernovae (SNe). The production of $^{23}$Na is quite substantial, with each star typically generating $\sim$ 0.1 $M_\odot$ of $^{23}$Na, prior to the onset of $^{20}$Ne burning.  Carbon burning can persist for times $\gtrsim 10^4$ years, during which the $^{23}$Na
is maintained at temperatures $\sim 10^9$K, optimal for axion production. There are several hundred Milky Way stars currently burning carbon and synthesizing $^{23}$Na. 

To determine the flux at Earth, one needs to address the relevant nuclear physics of axion production and propagation in the massive star from which it originates, then develop a galactic model to predict statistically the number, location, mass, and evolutionary phases of the contributing stellar sources.
\vspace*{.2cm}
\\
\textbf{{The axion-nucleus coupling:}}
The axion or axion-like particle (ALP) interaction with nucleons is
\begin{equation}
	\mathcal{L} = {1 \over 2M_N}  \bar{N}\gamma^\mu \gamma_5 (g^0_{aNN} +g^3_{aNN} \tau_3 ) N \, \partial_\mu a \,,
	\label{eq:ncoupling}
\end{equation}
where $M_N$ is the nucleon mass, $\tau_3$ is the third component of isospin,  $g_{aNN}^0 \equiv (g_{app}+g_{ann})/2$, and $g_{aNN}^3 \equiv (g_{app}-g_{ann})/2$. 

The nuclear transition probability can be obtained from 
Eq. (\ref{eq:ncoupling}) by equating the nuclear current to the sum over single-nucleon currents (after a nonrelativistic reduction to leading order in $1/M_N$), then performing a standard 
multipole decomposition to account for the momentum $q_a$ transferred
by the axion of energy $\epsilon_a$ to the nucleus.  One finds that axion absorption/emission is governed by two familiar electroweak operators,
the longitudinal spin $\Sigma_J^{\prime \prime}$ and axial charge $\Omega_J^\prime$ \cite{Donnelly:1980tsp}.  These abnormal-parity operators carry $J^\pi$ of $0^-$, $1^+$, $2^-$, $\ldots$.  
Details can be found in \cite{HLRR2}.

The low-energy $^{23}$Na transition can be treated in the
allowed limit, where only the $1^+$ multipole contributes. 
Following \cite{PhysRevLett.66.2557,PhysRevD.37.618}
(see details in \cite{HLRR2}), we normalize the axion emission rate to the known $\gamma$ decay rate of the
excited state, reducing the dependence on nuclear models. This yields
\begin{equation}
{\omega_a \over \omega_\gamma} = {1 \over 2 \pi \alpha} {1 \over 1+ \delta^2}\, {q_a^3 \over \epsilon_a^3} \, \left[ {g^0_{aNN}\beta + g^3_{aNN} \over (\mu_0-{1 \over 2}) \beta + \mu_1-\eta} \right]^2\,,
\end{equation}
where $\delta$ is the E2/M1 mixing ratio, and $\mu_0=0.88$ and $\mu_1=4.706$ are the isoscalar and isovector magnetic moments, respectively.
The nuclear physics is isolated in two matrix-element ratios of spin and orbital angular momentum operators
\begin{equation}
\beta \equiv  {
\langle J_f \|  \sum\limits_{i=1}^A \boldsymbol\sigma(i)  \| J_i \rangle
\over 
\langle J_f \| \sum\limits_{i=1}^A \boldsymbol\sigma(i) \,\tau_3(i) \| J_i \rangle 
},
~~
\eta \equiv - {
\langle J_f \| \sum\limits_{i=1}^A \boldsymbol\ell(i)\, \tau_3(i) \| J_i \rangle 
\over 
\langle J_f \| \sum\limits_{i=1}^A \boldsymbol\sigma(i)\, \tau_3(i) \| J_i \rangle 
}.
\end{equation}
For $^{23}$Na, the shell-model calculation of \cite{PhysRevLett.66.2557} gives $\beta=0.884$ and $\eta=-1.20$. The effective coupling to $^{23}$Na is
\begin{equation}
    g_{aNN}^0 \beta + g_{aNN}^3 \propto (g_{app}-0.062g_{ann}) \equiv  g_{aNN}^\mathrm{eff~^{23}Na}\,.
\end{equation}

The inverse process, resonant absorption of monochromatic axions
on $^{23}$Na, governs both detection in NaI and the stellar opacity for axions. 
The cross section can be expressed in terms of the spin-averaged and -summed transition prbability $|\mathcal{\bar{M}}|^2$ as
\begin{equation}
    \sigma(\epsilon_a) = \frac{\pi}{q_a} |\mathcal{\bar{M}}|^2 \, \delta(\epsilon_a-E_\mathrm{ex}).
    \label{eq:crosssection}
\end{equation}
The axions driving this transition are generated at 
stellar temperatures $\sim 10^9$K, where the line axions are
thermally broadened by the nuclear motion.  We write the flux
as the total $\bar{\phi}_a$ multiplied by a (normalized) thermal probability distribution,
\begin{align}
   & \phi_a(\epsilon_a) = \bar{\phi}_a \, P(\epsilon_a) = \frac{\bar{\phi}_a}{\sigma_\mathrm{TH} \sqrt{2\pi}}\, \exp\left[-\frac{(\epsilon_a-\bar{\epsilon}_a)^2 }{ 2\sigma_\mathrm{TH}^2}\right], \nonumber \\
   &~~ \sigma_\mathrm{TH}(T)  = \bar{\epsilon}_a \sqrt{ \frac{k_B T }{ M_T}}~\Rightarrow ~\sigma_\mathrm{TH}(T_9=1)  \approx 0.88 \, \mathrm{keV},
   \label{eq:thermal}
\end{align}
where $M_T$ is the target mass, $k_B$ is the Boltzmann constant, $\bar{\epsilon}_a \approx 440.2$ keV is the nuclear transition energy, and $T_9$ is the temperature
in units of $10^9$K. The line centroid can be shifted by two effects: the energy lost
to nuclear recoil and the gravitational red shift.  The first is
entirely negligible, while the second depends on progenitor 
properties, but in \cite{HLRR2} is found to be $\approx 150~\mathrm{eV} \ll \sigma_\mathrm{TH}$.
Consequently, we can take $P(\epsilon_a) \approx
P(\bar{\epsilon}_a) = 1/\sqrt{2 \pi} \, \sigma_\mathrm{TH}$,

We integrate Eq.~(\ref{eq:crosssection})
over the normalized line profile of Eq.~(\ref{eq:thermal}) to obtain the resonant cross section, again using the known radiative width of the excited state,
$\Gamma_\gamma \equiv \hbar \omega_\gamma$, as an experimental
normalization.  Keeping
track of the statistical factors $[J]\equiv \sqrt{2J+1}$, we find \cite{HLRR2}
\begin{equation}
    \langle \sigma \rangle =  \frac{{\pi} }{ \alpha (1+\delta^2)} \frac{\Gamma_\gamma }{ \epsilon_a^3} \frac{q_a }{ \sqrt{2\pi} \sigma_\mathrm{TH}} {[J_{ex}]^2 \over [J_{gs}]^2}\left[ \frac{g^0_{aNN}\beta +g^3_{aNN} }{ (\mu_0-\frac{1}{2}) \beta +\mu_1 -\eta} \right]^2
    \label{eq:resonant2}
\end{equation}
where $\sigma_\mathrm{TH}$ is evaluated for the temperature at the point of axion production.  The enhancement due to resonant absorption is captured
in the ratio $q_a/\sqrt{2 \pi}\sigma_\mathrm{TH}$. We use ENSDF values for
$\Gamma_\gamma = 4.01 \times 10^{-4}$ eV and $\delta=0.065$.

Equation~(\ref{eq:resonant2}) also governs the probability that an axion produced at some stellar
point $i$ is reabsorbed on $^{23}$Na at some second stellar point $f$, removing it from the flux.  The thermal
motions of the emitting and absorbing nuclei both contribute to the thermal line broadening, so that Eq.~(\ref{eq:thermal})
becomes
\begin{equation}
    \sigma_\mathrm{TH}(T) \rightarrow \sigma_\mathrm{TH}(T_i,T_f) = \bar{\epsilon}_a \sqrt{ \frac{k_B (T_i+T_f) }{ M_T}}\,.
    \label{eq:thermal2}
\end{equation}
This process dominates the opacity for 440 keV axions.
\vspace*{.2cm}
\\
\textbf{{Galactic modeling of the $^{23}$Na axion flux:}}  The yield of $^{23}$Na
axions can be computed for a given star from its temperature, density, and mass-fraction
profiles, which evolve over time. The total flux of $^{23}$Na axions can then
be computed from a galactic model that describes the spatial distribution,
masses, and evolutionary stages of today's carbon-burning stars.
The procedure requires one to evolve an ensemble of progenitors, which then are embedded in the galactic model, yielding the axion flux as a function of
its coupling. 
The galactic simulation must be repeated many times, as its predictions are
statistical, to determine the flux probability distribution

This program was carried out in \cite{HLMR}.  
An ensemble of contributing stars, the progenitors of ONeMg WDs and
electron-capture and core-collapse SNe, were evolved
using MESA \cite{Paxton2011}.  The galactic model was fitted to disk and bulge stellar population counts with an initial mass function of the Salpeter form. The adopted WD and SN galactic rates were 1/175y and 1/40y.

The resulting distribution for the axion flux at Earth is well descried by a two-sided Gaussian/Lorentzian
\begin{equation}
    P[\phi]=\sqrt{2 \over \pi } {1 \over \sigma_1+\sigma_2} \left\{ \begin{array}{cc} e^{-(\phi-\phi_0)^2/2 \sigma_1^2}, & \phi<\phi_0 \\ \left[ 1+\pi {(\phi-\phi_0)^2 \over 2 \sigma_2^2} \right]^{-1}, & \phi>\phi_0 \end{array} \right.
\end{equation}
where $\sigma_1 = 0.151 \phi_0$ and $\sigma_2=0.492 \phi_0$ with $\phi_0$ being the most probable flux, 
\begin{equation}
    \phi_0 = 0.173 \times 10^6 \left| { g_{aNN}^\mathrm{eff \, ^{23}Na} \over 10^{-7}} \right|^2\mathrm{cm^{-2}s^{-1}}.
\end{equation}
The skewed distribution reflects the underlying statistics.  Because there are many sources 
contributing at any given time, the low-flux side of this distribution is
a Gaussian with a relatively small standard deviation $\sigma_1$: a suppressed flux requires an unlikely correlation among many sources, forming a void around the Earth.  In contrast, the high-flux side is Lorentzian with a relatively large
$\sigma_2$:  a single progenitor 
anomalously close to Earth elevates the flux.

The Gaussian form of $P[\phi]$ for low $\phi$ is important for experiment, as it makes very low fluxes unlikely. In an experiment, the distribution $P[\phi]$ would be combined with detector counting uncertainties to exclude values of $|g_{aNN}^\mathrm{eff~^{23}Na}|$ at a desired confidence level.  For present exploratory purposes, we follow \cite{HLMR} in using a fixed representative flux, the median value
\begin{equation}
    \langle \phi_a^\mathrm{galactic} \rangle = 0.214 \times 10^6 \left| { g_{aNN}^\mathrm{eff \, ^{23}Na} \over 10^{-7}} \right|^2\mathrm{cm^{-2}s^{-1}} \,.
    \label{eq:standard}
\end{equation}
This value exceeds $\phi_0$ because the probability that today's flux lies in the Lorentzian portion of $P[\phi]$
is $\sigma_2/(\sigma_1+\sigma_2) \approx$ 0.76.

The possibility that today is unusual, a time of elevated axion flux, arises because of a recent claim that
Betelgeuse is burning carbon \cite{Saio}.  This conclusion, based
on an analysis of the star's brightness
variations, leads to an axion flux at Earth nearly an order of
magnitude larger than that of Eq.~(\ref{eq:standard}) \cite{HLRR2}.
The {\it a priori} probability of such a large flux is $\approx 2\%$. We will not consider this possibility further here.  See \cite{HLRR2} for discussions and alternative explanations of the brightness variations.
\vspace*{0.2cm}
\begin{figure*}[!t]
\centering
\includegraphics[scale=0.3]{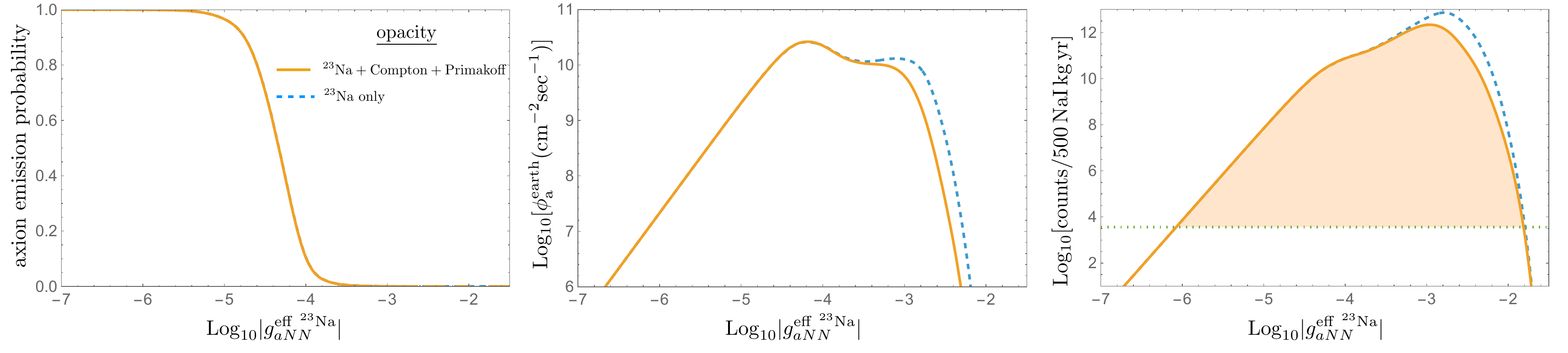}
\caption{The axion emission probability (left, calculated for an 11$M_\odot$ progenitor), $^{23}$Na axion flux (middle), and the axion counts in a NaI experiment (right) as functions of $|g_{aNN}^\mathrm{eff~^{23}Na}|$. The shaded region is the $3\sigma$ exclusion for a 500 kg-yr NaI exposure. The dashed lines (obscured by the solid line in the left panel) result if resonant reabsorption is the only opacity source.}
\label{fig:Count}
\end{figure*}
\\
\textbf{{Escape probability of $^{23}$Na axions:}}  Once a $^{23}$Na axion is produced, any interaction that reabsorbs the axion removes it from the flux.
An important advantage of $^{23}$Na axions is that their low energy prevents them from breaking up nuclei, the process that typically dominates 
axion opacities \cite{RAFFELT1982323}.  This makes resonant reabsorption on $^{23}$Na the only important nuclear process.  Other contributions 
come from Compton scattering off electrons and the Primakoff process. To access 
their importance, the axion couplings $g_{aee}$ and
$g_{a\gamma \gamma}$ must be related to $g_{aNN}^\mathrm{eff~^{23}Na}$ through
a model. We use
DFSZ axion couplings \cite{DINE1981199,osti_7063072} evaluated for $\sin^2\beta = 0.5$.
As detailed in \cite{HLRR2}, Primakoff absorption turns out to be unimportant,
while Compton scattering contributions are limited to a specific coupling window.
To the extent that DFSZ couplings are typical, a single coupling $g_{aNN}^{\mathrm{eff~^{23}Na}}$
controls $^{23}$Na axion creation, reabsorption, and detection.

The escape probability is given by 
\begin{equation}
    P_\mathrm{esc} [z_0]= \exp\left[-\int_{z_0}^{z_S}~\sum_i n_i(z) ~\sigma_i ~dz \right].
    \label{eq:prob}
\end{equation}
where $n_i$ is the number density of a given axion absorber and $\sigma_i$ is the cross section.
The integral is performed along the line connecting $z_0$, the point where the axion is produced, to 
the stellar surface, denoted as the point $z_S$.  The sum is performed over all channels $i$, folding the cross section with the target number density $n_i(z)$.

The calculation requires the line integral
to be performed for each coordinate $z_0$ within the star, weighted by the
probability that an axion is produced in an infinitesimal volume surrounding
that point. For larger $|g_{aNN}^\mathrm{eff~^{23}Na}|$, the mean free path of the axion becomes much smaller than the radius of the carbon-burning region.
In this case, the only axions that can avoid reabsorption on $^{23}$Na 
are emitted from the outside edge of the carbon-burning region, within the 
hemisphere facing Earth.  This ``axiosphere" becomes thinner with increasing
$|g_{aNN}^\mathrm{eff~^{23}Na}|$, eventually cutting off the flux. A careful integration over angles becomes important.

Results are given in Fig.~\ref{fig:Count}.  The left panel, computed for a 11$M_\odot$ star
1000 years prior to the end of carbon burning, shows
reabsorption becoming important for $|g_{aNN}^\mathrm{eff~^{23}Na}| \gtrsim 10^{-5}$.  While the calculation includes reabsorption on $^{23}$Na, Compton scattering, and the Primakoff
process, on a linear scale the result is indistinguishable from that obtained with
resonant reabsorption alone.

The middle panel gives the galactic $^{23}$Na flux at Earth. As the stellar production grows as $|g_{aNN}^\mathrm{eff}|^2$, a broad maximum in the flux forms between couplings $\sim 10^{-3}-10^{-5}$, with
$\phi_a \approx 10^{10}$/cm$^2$/sec. The peak flux is $\approx 2.1 \times 10^{10}$ axions/cm$^2$/sec.  The impact of Compton scattering is now visible but small, slightly diminishing the sensitivity to large couplings.

The right panel shows counts for a 500 kg-yr NaI exposure.  As the
detection introduces another factor of $|g_{aNN}^\mathrm{eff~^{23}Na}|^2$, the curves shift to the right while flattening at lower couplings. The $3\sigma$ excluded region
is based on the
estimate below that $\sim 10^4$ axion-induced events at 440 keV would stand
out above backgrounds.
\vspace*{.2cm}
\\ 
\textbf{{Detection sensitivities in NaI:}} 
NaI serves as both a source of 440 keV conversion $\gamma$'s and a detector for those $\gamma$'s. Here, we assume
an exposure of 500 kg-yr, equivalent to about two years of data in the 242 kg DAMA/LIBRA detector. (For reference, the collaboration's WIMP exposure
is nearing 3000 kg-yr.)  Such DM detectors are operating at high gain, focused on deposited energies  $\lesssim 10$ keV.  A search for a 440 keV signal would be conducted at low gain.
Some existing detectors, such as COSINE-100, can operate at high and low gains
simultaneously, which would allow parallel WIMP and axion searches \cite{reina}.

To estimate detection sensitivities, we compare background rates at 440 keV with axion counting rates.  The $3\sigma$ statistical uncertainty in the measured background counting rate in the vicinity 
of the 440 keV peak is
\begin{equation}
   \Delta_\mathrm{bg}^\mathrm{stat} \approx 3 \sqrt{M\, t\,  b\,  \Gamma_\mathrm{FWHM}}\,,
\end{equation}
where $b$ is the background rate per unit time, mass, and energy, $t$ is the integration time,
$M$ is the detector mass, and $\Gamma_\mathrm{FWHM}$ is the detector's resolution. We take $Mt$ = 500 kg-yr; $\Gamma_\mathrm{FWHM} \approx$ 22 keV,
from the COSINE result that $\Delta E/E \approx 0.05$
above 100 keV \cite{COSINEBG} and 
$b$ = 0.37 counts/kg/d/keV, obtained from averaging over
five COSINE-100 crystals in the 430-440 and 440-450 keV bins \cite{bins}.
This yields $\Delta_\mathrm{bg}^\mathrm{stat} \approx 3660$ events. 

We extract a $3\sigma$ exclusion for $|g_{aNN}^\mathrm{eff~^{23}Na}|$ by demanding that the counting rate not exceed $\Delta_\mathrm{bg}^\mathrm{stat}$, yielding
\begin{equation}
    8.4 \times 10^{-7} \lesssim |g_{aNN}^\mathrm{eff~^{23}Na}| \lesssim 1.5 \times 10^{-2}\,,
\end{equation}
as shown in the right panel of Fig.~\ref{fig:Count}. In an experiment, we anticipate that this estimate would be replaced by a maximum likelihood analysis, using a background model fitted to background rates in bins surrounding 440 keV.
\begin{figure*}[!t]
\centering
\includegraphics[scale=0.27]{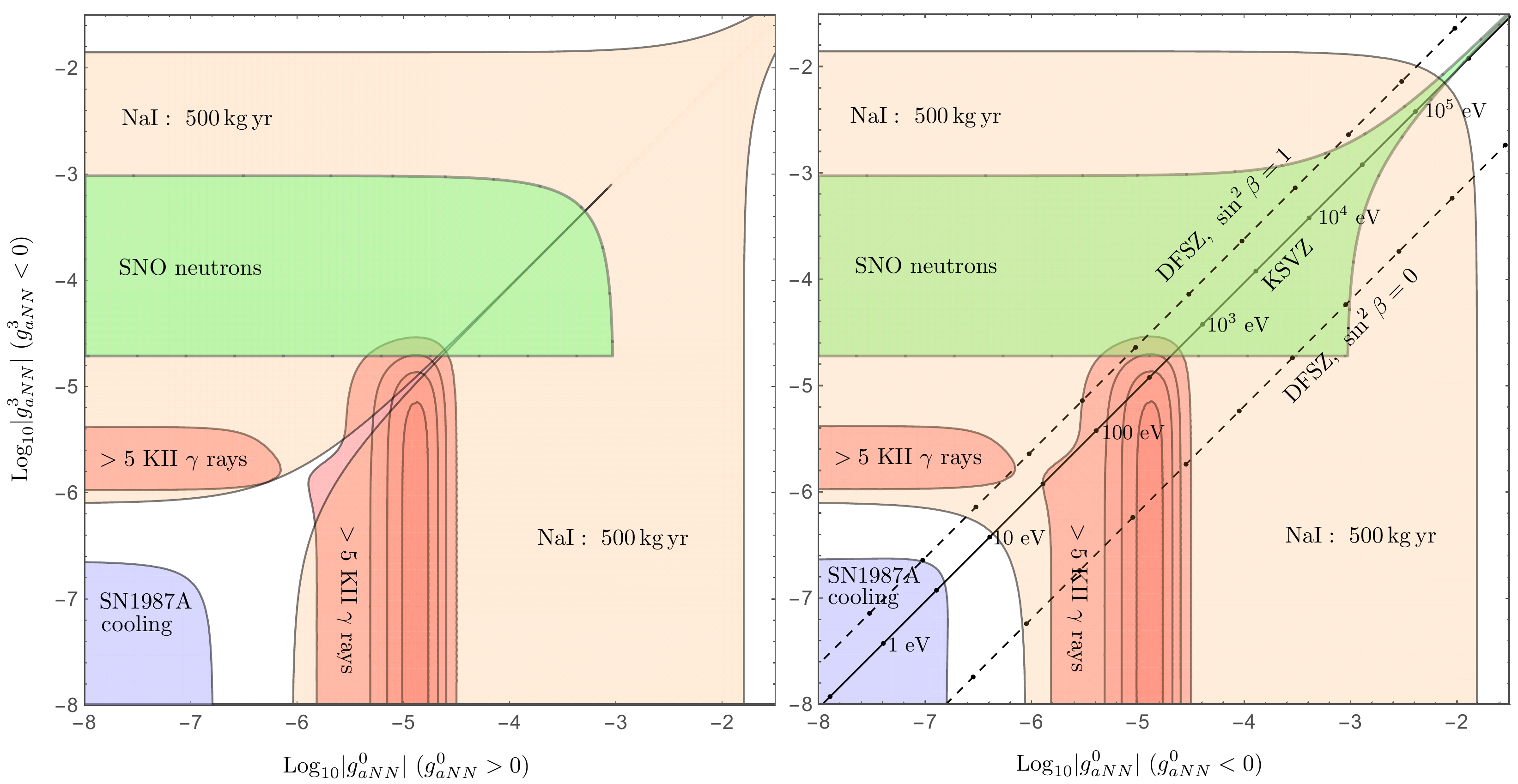}
\caption{Potential exclusions on axion-nucleon couplings assuming a 500\,kg-y NaI experiment (tan shaded). Existing astrophysical limits come from SNO (green), SN1987A KII $\gamma$ (red) and SN1987 cooling (blue). The SN1987A KII $\gamma$ exclusion contours correspond to 5, 15, 25, and 35 counts above background.}
\label{fig:limits}
\end{figure*}
\vspace*{.2cm}
\\
\textbf{{Constraints on Turner-window axions:}}
The ``Turner window" refers to axions with couplings sufficiently large to evade the SN1987A cooling bound, typically with $m_a \gtrsim$ 1 eV \cite{TURNER199067}.  Most of the relevant constraints on such axions come from astrophysical observations that probe a specific isospin combination of couplings.  We assume an ALP with couplings $g^0_{aNN}$ 
and $g^3_{aNN}$, without specifying the relationship to the ALP mass. If
a model is adopted, such as the DFSZ \cite{DINE1981199,osti_7063072} or KSVZ \cite{PhysRevLett.43.103,SHIFMAN1980493} axion, $m_a$ becomes a function
of $g_{aNN}^0$ and $g_{aNN}^3$, defining a trajectory or region in the 
$g_{aNN}^0-g_{aNN}^3$ plane, as in Fig.~\ref{fig:limits}.

The constraints that future NaI experiments might establish are shown in Fig.~\ref{fig:limits} along with three astrophysical constraints discussed
in detail in \cite{HLRR2}, but only summarized here: \\
\noindent
1. Limits derived from the production of neutrons in the SNO detector, from solar axions breaking up deuterium \cite{PhysRevLett.126.091601}. The solar source are the 5.5 MeV axions from $p+d \rightarrow {^3 \mathrm{He}}+ a$,
as first discussed in \cite{RAFFELT1982323}. The analysis of \cite{HLRR2}
largely supports that of \cite{PhysRevLett.126.091601}, though with a revised
upper bound
\begin{equation}
1.9 \times 10^{-5} < |g_{aNN}^3|~~~  \mathrm{and}~~     |g^3_{aNN}-g^0_{aNN}|< 0.95 \times 10^{-3}
\label{eq:SNO}\,.
\end{equation}
The upper bound comes from the observation of \cite{RAFFELT1982323} that the odd-neutron isotopes $^{17}$O and $^{13}$C dominate the solar opacity for 5.5 MeV axions due to their low breakup thresholds.\\
\noindent
2. Constraints derived from photo-deexcitation of excited states of
$^{16}$O in the Kamiokanda II (KII) detector, following absorption of axions
produced by SN1987A.  The limits obtained in \cite{PhysRevLett.65.960}, ruling out couplings $g_{aNN}$ from $\approx 10^{-6}$ to
$ \approx 10^{-4}$, were based on the assumption that axions produced in
bremsstrahlung exit the star once the reabsorption process
$aNN \rightarrow NN$ becomes inefficient.  This absorption scales with density as $\rho^{5/3}$, as
it requires correlated nucleons. However, absorption on nuclei --- to bound states and to breakup channels ---  scales as $\rho^{2/3}$ and thus can
be important over a more extended region of the star.  This additional absorption
was evaluated in \cite{HLRR2} using a SN1987A explosion model provided by the
Garching group \cite{Garching,Garching2}.  Absorption on iron group elements and $^{16}$O is very effective
in removing lower-energy axions.  In the calculations of
\cite{PhysRevLett.65.960}, the strongest source of KII $\gamma$'s came from
de-excitation of the quasi-bound 10.96 MeV $0^-0$ state in $^{16}$O.  
Before reaching KII, an axion must pass through the star's oxygen layer, 
where resonant absorption creates a deep minimum at the same 10.96 MeV energy.
Similarly, axions with energies $\gtrsim 19$ MeV are absorbed by freezeout 
$\alpha$'s, the most abundant absorber in the layers just outside the axiosphere:
0.3 sec after core bounce, $\alpha$'s dominate the abundance by
$r \approx 3 \times 10^7$cm.  In \cite{HLRR2} these effects are fully evaluated, leading to the much reduced exclusion area in Fig.~\ref{fig:limits}.\\
3. Limits obtained from axion cooling of SN1987A, which 
would shorten the time $\tau_\nu \approx 3$ sec over which burst neutrinos were observed \cite{PhysRevLett.58.1490,PhysRevLett.58.1494}. For axions to
compete with neutrinos in SN cooling, they
must have couplings to matter that are comparable to or weaker than 
the weak scale, so they can promptly escape. The argument is indirect,
and the unexpected 12.5 sec duration of the SN1987A neutrino burst has
generated some concerns \cite{PhysRevD.101.123025,Caputo:2024oqc}.  The constraint plotted in Fig.~\ref{fig:limits},
\begin{equation}
   1.19 \times 10^{-9} < [(g_{app}+0.435 g_{ann})^2 +1.45 g_{ann}^2]^{1/2} \lesssim 3 \times 10^{-7}\,,
    \label{eq:cooling}
\end{equation}
is from \cite{Caputo:2024oqc}.  The upper bound is the trapping limit.
\vspace*{.2cm}
\\
\textbf{{Potential impact of NaI searches:}} 
Figure \ref{fig:limits} gives the results for ALPs with dominant proton
(right panel, same sign $g^0_{aNN}$ and $g^3_{aNN}$)  and neutron
(left, opposite sign) couplings.  The right panel includes the KSVZ axion track,
parameterized by the axion mass as indicated, and the 
boundaries of the DFSZ area, defined by $\sin^2{\beta}=0$ and 1.
Note that if one considers a particular axion model, then additional constraints
can arise from limits on $g_{a\gamma \gamma}$ and $g_{aee}$.  Also note that
the limits assume $m_a \ll \epsilon_a$, which in most models would typically hold
within the various excluded regions.

It is clear that a NaI experiment with a 500 kg-yr dataset would have impact. The excluded region (tan shaded) extends over most of the plane, encompassing both the SNO neutron and KII photon regions.
The experiment is particularly clean because axion production, 
propagation in the star, and detection are governed by a single 
coupling, $|g_{aNN}^\mathrm{eff~^{23}Na}|$. The axion signal is
continuous and distinctive, not easily mimicked by backgrounds.

We contrast this with constraints derived from
SN1987A, a unique and perhaps atypical event, as the 11 KII neutrino events were spread irregularly over
$\sim$ 13 sec.  Given the KII background rate of $\approx$ 0.6 Hz, it is not clear what rate of KII $\gamma$s should be considered a 3$\sigma$ event.  For this reason we show event contours in Fig. \ref{fig:limits}.

The galaxy's flux of $^{23}$Na axions creates an opportunity to derive additional value from past
investments in NaI WIMP detectors.  Resonant
absorption has already been exploited for $^{57}$Fe axions.
The application to $^{23}$Na strikes us as much more attractive: NaI is both the $\gamma$
source and detector -- one is not limited to thin targets; the $^{23}$Na isotopic abundance is 100\%, not 2\%; and the
unpaired nucleon is a proton, to which QCD axions are more strongly coupled.

The main goal of this letter is to bring the possibility of 440 keV resonant-absorption axion experiments to the attention of the experimental collaborations who have developed NaI detectors. As we have noted, some detectors may allow simultaneous searches for light WIMPs and 440 keV axions. Regardless, their use in axion searches creates a significant new discovery opportunity.
\section*{Acknowledgments}
We are very grateful to Thomas Janka, Daniel Kresse, and Annop Wongwathanarat for making available output from the Garching group's SN1987A simulations, and to Reina Maruyama, Gyunho Yo, Eunju Jeon, and Karlheinz Langanke for helpful discussions.  WH acknowledges support by the US Department of Energy under grants DE-SC0004658, DE-SC0023663, and DE-AC02-05CH11231, the National Science Foundation under cooperative agreement 2020275, and the Heising-Simons Foundation under award 00F1C. This work was carried out under the auspices of the National Nuclear Security Administration of the U.S. Department of Energy at Los Alamos National Laboratory under Contract No. 89233218CNA000001. ER acknowledges support by the Laboratory Directed Research and Development program of Los Alamos National Laboratory under project numbers 20251163PRD3 and 20260043DR. AR acknowledges support from the Natural Sciences and Engineering Research Council of Canada (NSERC) and  Arthur B. McDonald Canadian Astroparticle Physics Research Institute. Research at Perimeter Institute is supported by the Government of Canada through the Department of Innovation, Science, and Economic Development, and by the Province of Ontario.

\bibliographystyle{apsrev4-1}
\bibliography{Na23_new.bib}
\end{document}